\documentclass[a4paper,12pt,fleqn]{cas-sc}

\usepackage[authoryear,longnamesfirst]{natbib}
\usepackage{acronym}

\usepackage{xcolor}
\usepackage{graphicx} 
\usepackage{subcaption}
\usepackage[normalem]{ulem}
\usepackage{amsmath}
\usepackage{amssymb}

\def\tsc#1{\csdef{#1}{\textsc{\lowercase{#1}}\xspace}}
\tsc{WGM}
\tsc{QE}

\def\b0{\mbox{\boldmath $0$}}

\def\btheta{\mbox{\boldmath $\theta$}}
\def\beps{\mbox{\boldmath $\epsilon$}}
\def\bvareps{\mbox{\boldmath $\varepsilon$}}
\def\bphi{\mbox{\boldmath $\phi$}}
\def\bxi{\mbox{\boldmath $\xi$}}
\def\bzeta{\mbox{\boldmath $\zeta$}}

\def\bH{\mbox{\bf H}}

\def\bK{\mbox{\bf K}}

\def\bP{\mbox{\bf P}}

\def\bd{\mbox{\bf d}}
\def\bff{\mbox{\bf f}}
\def\bg{\mbox{\bf g}}

\def\bu{\mbox{\bf u}}

\def\bx{\mbox{\bf x}}

\newlength{\figwidthfull}
\newlength{\figwidthhalf}
\newlength{\figspace}
\renewcommand{\exp}[1]{{\rm exp}\ensuremath{\left [{#1} \right]}}

\newcommand{\beq}{\begin{equation}}
\newcommand{\eeq}{\end{equation}}

\acrodef{EKF}{Extended Kalman Filter}
\acrodef{KF}{Kalman Filter}
\acrodef{EnKF}{Ensemble Kalman Filter}
\acrodef{pdf}{probability density function}
\acrodef{PF}{Particle Filter}
\acrodef{WEC}{wave energy converter}

\begin{document}
\let\WriteBookmarks\relax
\def\floatpagepagefraction{1}
\def\textpagefraction{.001}

\shorttitle{Wave Sequential Data Assimilation in Support of Wave Energy Converter Power Prediction}    

\shortauthors{M. Khalil, C. Michelén, K. Raghukumar, A. Dallman}  

\title [mode = title]{Wave Sequential Data Assimilation in Support of Wave Energy Converter Power Prediction}

%

\author[1]{Mohammad Khalil}[type=editor,
                        orcid=0000-0001-6579-7127]

\ead{mkhalil@sandia.gov}

\author[2]{Carlos Michelén~Ströfer}

\author[3]{Kaustubha Raghukumar}

\author[2]{Ann Dallman}

\affiliation[1]{organization={Sandia National Laboratories},
                city={Livermore},
                state={California},
                postcode={94551}, 
                country={USA}}
                
\affiliation[2]{organization={Sandia National Laboratories},
                city={Albuquerque},
                state={New Mexico},
                postcode={87185}, 
                country={USA}}
                
\affiliation[3]{organization={Integral Consulting Inc.},
                city={Santa cruz},
                state={California},
                postcode={95060}, 
                country={USA}}


\begin{abstract}
Integration of renewable power sources into grids remains an active research and development area, particularly for less developed renewable energy technologies such as wave energy converters (WECs). WECs are projected to have strong early market penetration for remote communities, which serve as natural microgrids. Hence, accurate wave predictions to manage the interactions of a WEC array with microgrids is especially important. Recently developed, low-cost wave measurement buoys allow for operational assimilation of wave data at remote, site specific locations where real-time data have previously been unavailable. 

We present the development and assessment of a wave modeling framework with real-time data assimilation capabilities for WEC power prediction. The availability of real-time wave spectra from low-cost wave measurement buoys allows for operational data assimilation with the ensemble Kalman filter technique within a hybrid modeling procedure whereby physics-based numerical wave models are combined with data-driven error models that aim to capture the discrepancy in prescribed boundary conditions. With that aim, measured wave spectra are assimilated for combined state and parameter estimation while taking into account model and observational errors. The analysis allows for more accurate and precise wave characteristic predictions at the locations of interest. Initial deployment data obtained offshore Yakutat, Alaska, indicated that measured wave data from one buoy that were assimilated into the wave modeling framework resulted in improved forecast skill in comparison to traditional numerical forecasts.
\end{abstract}


\begin{highlights}
\item Developed a data assimilation framework that leverages numerical spectral wave models for forecasting and the ensemble Kalman filter for data assimilation of near-shore wave spectra
\item Developed data-driven error models that aim to capture the discrepancy in prescribed boundary conditions of wave spectra
\item Assimilated real data obtained offshore Yakutat, Alaska, from low-cost wave measurement buoys, resulting in improved forecast skill in comparison to traditional numerical forecasts.
\end{highlights}


\maketitle


\section{Introduction}\label{sec:intro}

A number of different \ac{WEC} technologies are currently being developed to harness this largely untapped energy resource; however successful incorporation of wave energy projects to electric grids requires (1) forecasting available energy, (2) efficiently harnessing that energy (via multiple devices in a “farm” or array), and (3) modular management of sources, loads, and storage. Integration of renewable power sources into electrical grids remains an active research and development area, particularly for less developed renewable energy technologies such as \acp{WEC}. Effective grid management will require accurate and precise forecasts, particularly for high penetrations of renewables onto microgrids. Typical wave sea state forecasting has significant errors, which will compound when coupled with array power output. Further, understanding uncertainty of the power forecast (i.e., the forecast model for wave power produced by a \ac{WEC} array with and without improvements to incoming wave sea state forecasting) is critical to quantify and reduce. In this investigation, we provide a framework that couples data assimilation with wave forecasting in order to arrive at more accurate and certain wave forecasts. These forecasts will enable future efforts to balance energy variability on electrical grids to obtain energy resiliency and security, and to ensure that the potential of ocean wave energy is realized. This is enabled by the availability of low-cost wave measurement buoys. 

Recently, a data assimilation framework utilized SWAN to predict wave characteristics in the basin of the Black Sea [\cite{rusu2016multi}] and the Portuguese coastal area [\cite{guedes2011operational}]. Both of those approaches rely on extrapolating the difference between the measured data and simulation results and applying corrections at the boundaries of the computational domain. In this investigation, we will update the boundary conditions using state-of-the-art data assimilation strategies. Furthermore, our corrections will be non-linear-in-time versus ones that are linear-in-time as applied in [\cite{rusu2016multi,guedes2011operational}].

In the context of sequential data assimilation for real-time applications, \ac{KF} has been widely applied to estimate the statistical moments of the system state vector given noisy observational data [\cite{kalman60}], providing optimal estimates for such statistics in the case of linear systems with additive Gaussian noise. For nonlinear systems, Monte Carlo based sequential filtering algorithms have gained popularity due to their wide scope of applicability. These sampling-based methods represent the \ac{pdf} of the state vector using a finite number of possible states or trajectories. The two most popular classes of Monte Carlo based filters are the \ac{EnKF} [\cite{evensen06}] and \ac{PF} [\cite{tanizaki96,doucet00,ristic04}]. \ac{EnKF} and \ac{PF} can be easily applied to complicated black-box models with no access to derivative information (as is the case for the SWAN numerical wave model).

In addition to estimating the unknown state from noisy observations, we will attempt to correct pre-specified boundary conditions using time-varying error models, whose unknown parameters will also be inferred from the available data. One popular approach to solving the parameter estimation problem within sequential data assimilation settings involves augmenting the state vector with the unknown system parameters. The new augmented state vector is estimated using \ac{EnKF} in the same manner that the original state vector is estimated. The parameter space is only explored at the initialization of the algorithm according to the prior \ac{pdf} of the unknown parameters [\cite{andrieu03}], with performance tightly related to the choice of priors. For poorly selected prior, the filtering algorithms fail after a few assimilation cycles, a phenomenon commonly known as filter divergence. Unfortunately, in our setting the parameters belong to a purely data-driven model, making it more difficult to provide a sufficiently compact prior \ac{pdf} ahead of the assimilation experiment. To alleviate this problem, the parameters are perturbed by artificial noise in time to achieve convergence of the filtering algorithm [\cite{kitagawa98,khalil07_2}].

The paper is organized as follows. In Section 2, we provide an overview of joint state and parameter inference of dynamical systems using the ensemble Kalman filter for sequential data assimilation. In Section 3, we present an application of the novel data assimilation workflows to a case study involving an area offshore of Yakutat, Alaska, with results and discussions provided in Section 4. We conclude, in Section 5, with a summary of our findings and some additional observations in the context of future work.

\section{Sequential Data Assimilation}

In this section, we provide an overview of the ensemble Kalman filter (EnKF) as a sequential data assimilation technique for the joint inference of unknown state and parameters of dynamical systems.

\subsection{State-Space Representation of Discrete Dynamical Systems}
\label{dynamics}

In a data assimilation setting, the model and measurement equations for a discrete state-space representation of a (nonlinear) dynamical system are given by
[\cite{evensen2009data}]
\begin{align}
\label{par1}
\bu_{k+1} & = \bg \left( \bu_{k}, \bff_k, \btheta, \bvareps_k \right) \ , \\
\label{par2} \bd_j & = \bH \ \bu_{k(j)} +  \beps_{j} \ .
\end{align}
Here $\bu_k$ is the unknown state vector at time instance $t = t_k$, $\bg$ is the discrete vector-valued nonlinear model operator, $\bphi$ is the vector of unknown (or weakly known) modeling parameters, $\bff$ is a vector of deterministic (i.e. known) inputs, $\bvareps_{k}$ is a random vector to capture modeling uncertainties, and $\bd$, the measurement vector, relates to the true state through the linear measurement operator $\bH$ with $\beps_{j}$ being a random vector that captures the additive observational errors/uncertainties. Note that this can be extended to handle non-linear measurement operators (see [\cite{evensen2009data}] for details). The indices $k$ and $k(j)$
denotes the time steps for state evolution and arrival of measurement data, respectively.

\subsection{Joint State and Parameter Estimation}
In the framework of nonlinear filtering, one popular strategy to infer the unknown parameter vector $\btheta$ along with the state is to treat the parameters as a time-varying quantities, artificially perturbed randomly to avoid filter divergence (e.g. [\cite{chui99,khalil07_2}]). In this case, the state space model can be written as
\begin{align}
\label{par11}
\bu_{k+1} & = \bg \left( \bu_{k}, \bff_k, \btheta_k, \bvareps_k \right) \ , \\
& \btheta_{k+1} = \btheta_{k} + \bxi_k \ , \\
\end{align}
where $\bxi$ is usually chosen to be a multivariate normal random vector that artificially inflates the variance (uncertainty) of the unknown parameter vector to avoid filter divergence. The {\it augmented} state vector is formed
by appending the unknown parameter vector to the original state vector leading to the following augmented state space model:

\begin{align}
\label{par111}
& \bx_{k+1} = \mathbb{g} \left( \bx_{k}, \bff_k, \bzeta_k \right), \\
\label{par222} & \bd_j = \mathbb{H} \ \bx_{k(j)} +  \beps_{j}.
\end{align}

where

\begin{equation}
\bx_{k} = 
\begin{Bmatrix}
\bu_{k}\\
\btheta_{k}\\
\end{Bmatrix}
;\,\,\,\,\\
\bzeta_{k} = 
\begin{Bmatrix}
\bvareps_{k}\\
\bxi_{k}\\
\end{Bmatrix}
\end{equation}

Nonlinear filter can provide estimates of the conditional distribution of the uncertain state $\bu$ and parameters $\btheta$ given noisy observations $\bd$. In this investigation, \ac{EnKF} will be used to provide those estimates. Next, we will briefly present \ac{EnKF} for data assimilation in dynamical systems.

\subsection{Ensemble Kalman filter}
\label{sec:EnKF}

In \ac{EnKF}, first introduced by Evensen [\cite{evensen1994sequential}] as an alternative to the extended Kalman filter for nonlinear dynamical systems, a finite number
of Monte Carlo samples of the (augmented) state vector $\bx_k$ are propagated forward in
time using the original model operator, treated as a black box model. One main concept behind the success of \ac{EnKF} is that the statistical moments of the unknown state vector can be approximated using ensemble averaging. \ac{EnKF} also performs a linear {\it analysis} or update step that is inherited from \ac{KF}, essentially relying on second order moments (i.e. covariances). The linear analysis step offers
computational efficiency but introduces errors in the estimated conditional \ac{pdf} of $\bx_k$ since it ignored higher order moments in the target state \ac{pdf}.

In one iterations of data assimilation using \ac{EnKF}, we start with a prior \ac{pdf} of $\bx$ given by $\bx_k \sim p(\bx_k^f)$. This prior \ac{pdf} is the \ac{pdf} of the state at time $t = t_k$ prior to assimilating the available data that is available at that time instance. This prior \ac{pdf} is approximated using a set of Monte Carlo samples or trajectories of the state vector as provided through time-integration of the model operator. In the analysis step, \ac{EnKF} estimates the conditional pdf $\bx_{k(j)} \sim p(\bx_{k(j)}^a)$ given the available vector of observations $\bd_j$.

In summary, given model and measurement equations, as in Eqs.~(\ref{par111})-(\ref{par222}), \ac{EnKF} provides Monte Carlo samples of the state vector {\it approximately} distributed according to the conditional \ac{pdf} the state vector given available observations as follows [\cite{evensen06}]. Start by constructing an initial ensemble $\left\{ \bx_{0,i}^a \right\}$ of size $N$ with $i =
1,\hdots,N$,
using the prior \ac{pdf} of $\bx_0$ (prior to assimilating any data/observations). This prior \ac{pdf} relates to the knowledge that is available relating to the initial conditions of the dynamical system. For each subsequent iterations, perform the following three steps
\begin{enumerate}
\item Forecast step: Repeat the time-integration
\begin{align}
\bx_{k,i}^f & = \mathbb{g} \left( \bx_{k-1,i}^a, \bff_k, \bzeta_{k-1,i} \right)  \ , \ \ \ \ \ \ \ \ i = 1, ... , N
\end{align}
until observations are available for assimilation, where $\bzeta_{k-1,i}$ are Monte Carlo samples of the random vector $\bzeta_{k-1}$.

\item Analysis step:
\begin{align}
\bd_{j,i} & = \mathbb{H} \ \bx^f_{k(j),i} +  \beps_{j,i} \ , \ \ \ \ \ \ \ \ i = 1, ... , N \ \ \ \ \ \ \ \ \ {\rm ({\it perturbed} \ observations)} \\
\overline{\bd}_j & = \frac{1}{N} \sum_{i=1}^N \bd_{j,i}\,\,, \\
\overline{\bx}^f_{k(j)} & = \frac{1}{N} \sum_{i=1}^N \bx^f_{k(j),i}\,\,, \\
\bP_{xd} & = \frac{1}{N-1} \sum_{i=1}^N \left(\bx^f_{k(j),i} - \overline{\bx}_{k(j)}^f \right)\left(\bd_{j,i} - \overline{\bd}_{j} \right)^{\it T}, \label{x_cov_app}\\
\bP_{dd} & = \frac{1}{N-1} \sum_{i=1}^N
\left(\bd_{j,i} - \overline{\bd}_{j} \right) \left(\bd_{j,i} - \overline{\bd}_{j} \right)^{\it T}, \\
\bK_{k(j)} & = \bP_{xd} \bP_{dd}^{-1},\\
\bx_{k(j),i}^a  & = \bx_{k(j),i}^f + \bK_{k(j)} \left( \bd_{j,i} - \mathbb{H} \ \bx^f_{k(j),i} \right)  \ , \ \ \ \ \ \ \ \ i = 1, ... , N \ .
\end{align}

\end{enumerate}

In general, the number of trajectories $N$ is chosen to accurately capture the cross-correlation between the state forecasts and perturbed observations, as approximated by $\bP_{xd}$ in Eq.~(\ref{x_cov_app}).

\section{Application to Spectral Wave Modeling}

In this section, we describe an application of the data assimilation framework to a case study involving an area offshore of Yakutat, Alaska. For data assimilation purposes, spoondrift wave measurement buoys were deployed off the coast of Yakutat, providing the observations to be fused with computational model forecasts as provided by SWAN numerical wave model [\cite{booij1999third}].

\subsection{Case Study}

The data assimilation framework is applied to a case study involving an area offshore of Yakutat, Alaska, where the community is reliant on 100\% diesel energy generation. The community, however, is considering utilizing renewable electricity generation, in particular wave energy. The University of Alaska Fairbanks (UAF) has completed a resource assessment using measurements and numerical modeling [\cite{tschetter2016yakutat}]. A regional wave model was used in order to characterize the regional wave climate, and identify promising installation sites for \acp{WEC}. It was found that the average annual wave energy at the site is around 20 kW/m, and could provide more electricity than Yakutat’s electrical demand.

\begin{figure}[!ht]
    \centering
    \includegraphics[width=0.75\columnwidth]{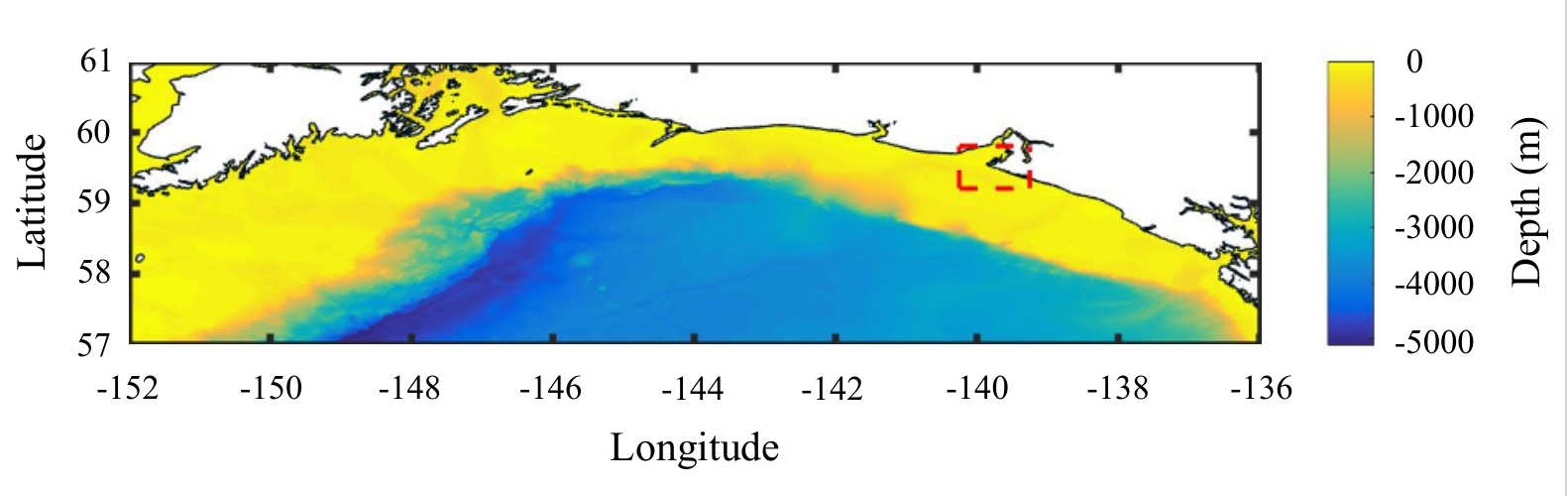}
    \caption{Coarse SWAN domain bathymetry.  The outline of the intermediate SWAN domain is indicated by the red dotted line.}
    \label{fig:domain_out}
\end{figure}

\begin{figure}[!ht]
    \centering
    \includegraphics[width=0.7\columnwidth]{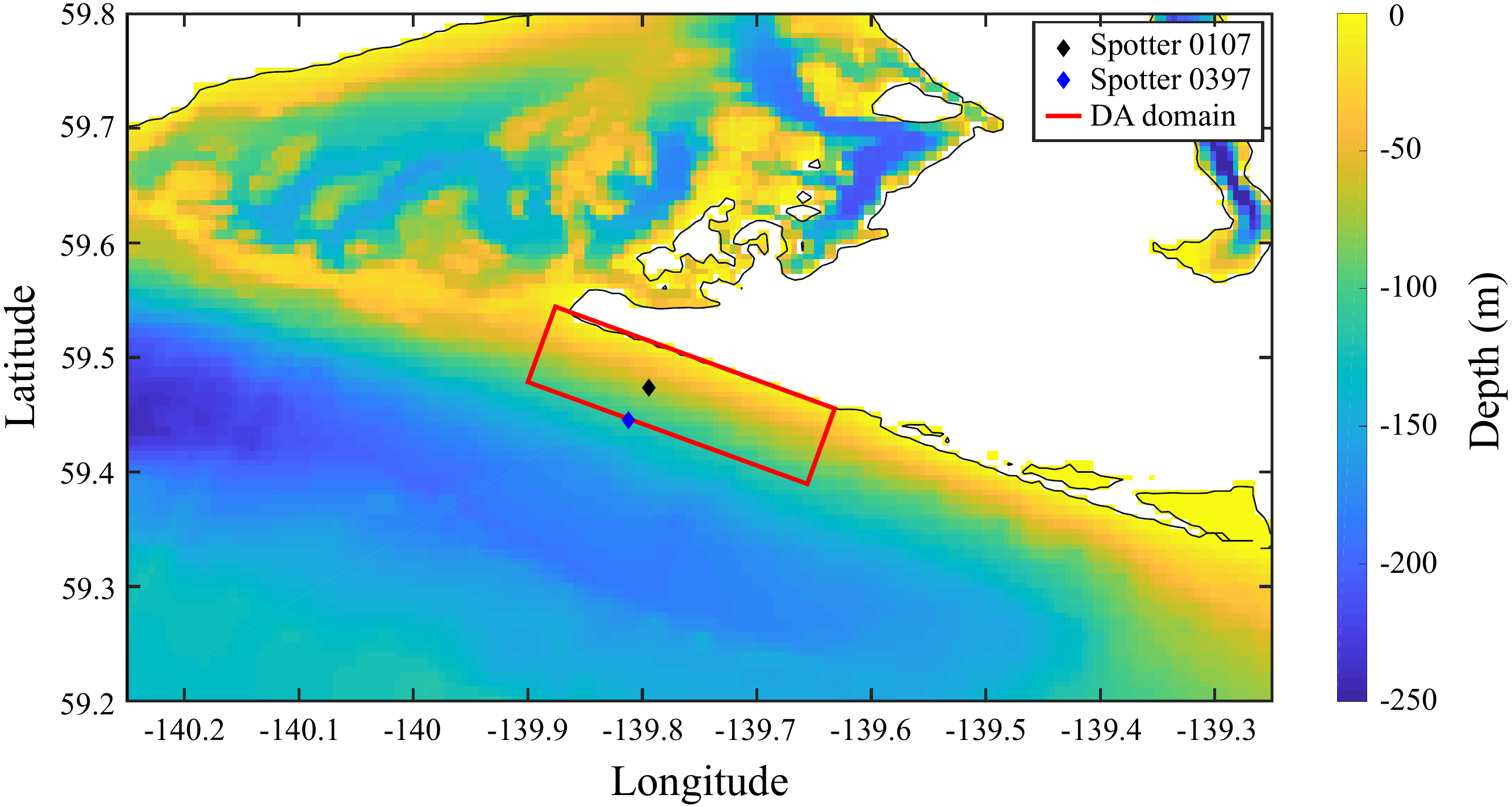}
    \caption{Intermediate SWAN domain showing the domain used in data assimilation and the location of the buoys providing observational data.}
    \label{fig:domain_in}
\end{figure}

The coastal geography and offshore bathymetry in the area of interest are shown in Fig.~\ref{fig:domain_out}. The baseline model setup for this analysis is taken from the UAF study [\cite{tschetter2016yakutat}], which included a 10-year wave model hindcast. The coarse model domain is shown in Fig.~\ref{fig:domain_out} and the intermediate model domain is shown in Fig.~\ref{fig:domain_in}. Additionally, the nearshore domain setup for data assimilation is shown in Fig.~\ref{fig:domain_in}, which is rotated by 20 degrees to be approximately parallel to the shore.

Offshore of Yakutat, deployment operations limited the deeper Spotter buoy (SPOT-0397) deployment location to about 3 nautical miles offshore, which in turn determined the extent of this application's model domain considered for data assimilation. The two Spotter buoys that were deployed in 2020 (at 67m and 109m depths) are also shown in Fig.~\ref{fig:domain_in}.

\subsection{SWAN Numerical Wave Model}

The SWAN model [\cite{booij1999third}] is a fully discrete spectral wave model which integrates the energy balance equation, describing the evolution of the wave spectrum in time, geographical and spectral spaces (defined in SWAN by the relative radian frequency  and the wave direction). The action density $N\left( \sigma, \theta \right)$ is equal to the energy density $E\left( \sigma, \theta \right)$ divided by the relative frequency $\sigma$, where $\theta$ is the wave direction. The evolution of the action density $N$ is governed by [\cite{komen1996dynamics}]:
\begin{equation}\label{eq:action}
\frac{\partial N}{\partial t} + \nabla_{\vec{x}} \cdot \left[ \left( \vec{c}_g + \vec{U} \right) N \right] + \frac{\partial c_\theta N}{\partial \theta} + \frac{\partial c_\sigma N}{\partial \sigma} = \frac{S_{tot}}{\sigma} \ ,
\end{equation}
where $N (\sigma,\theta)$ is the action density, $x$ is space, $t$ is time, $c_g$ represents propagation velocity of the action in geographical space, $C_\sigma$ is propagation velocity in frequency space, $c_\theta$ is propagation velocity in wave directional-space. The source term, $S_{tot}$, in SWAN captures various sources of excitation and external effects, including the atmospheric (e.g. wind) input, whitecapping dissipation, and decay due to bottom friction. The integration of the action balance equation, Eq.~(\ref{eq:action}), is accomplished with finite difference discretization in time, physical and spectral spaces. The SWAN model utilizes fully implicit numerical schemes.

\subsection{Available data}

For the data assimilation experiments, the available observations are in the form of two spectral bulk parameters, namely significant wave height and energy period, from the SPOT-0397 buoy near the border of the computational domain (see Fig.~\ref{fig:domain_in}). The significant wave height, $H_s$, and energy period, $T_e$, relate to the wave energy density as in
\begin{align}
H_s & = 4 \sqrt{m_0} \ ,\\
T_e & = \frac{m_{-1}}{m_0}\ ,
\end{align}
with the associated relevant spectral moments
\begin{align}
m_0 & = \int E \left( \omega, \theta \right) {\rm d} \omega \, {\rm d} \theta \ ,\\
m_{-1} & = \int \omega^{-1} E \left( \omega, \theta \right) {\rm d} \omega \, {\rm d} \theta \ ,
\end{align}
at a particular time and spatial coordinate.

In the update step of EnKF, wave spectra over the domain were updated based on the difference in the two bulk parameters between the model predictions and the buoy data at the buoy location. The updated bulk parameters are subsequently used to update the descretized spectra using frequency and amplitude modulation as described in App.~\ref{app:spec_update}. The data assimilation framework integrates the wave energy density in time to the nearest 10th minute forecast interval at which the next data is available for assimilation, and incorporates the data when updating the state and four parameters that control the BC correction.

\subsection{Boundary Condition Correction}

Our data assimilation framework attempts to not only update the unknown state vector (discretized wave spectra, or summary statistics thereof, over the computational domain), but also correct the boundary conditions provided by the larger-scale model simulations (nested SWAN runs over increasingly larger computational domain, shown in Fig.~\ref{fig:domain_out} and Fig.~\ref{fig:domain_in}).

For the source of uncertainty in our computational model, we implemented a stochastic model for the boundary conditions which guarantees a certain level of uncertainty in the wave spectra along the boundary. This is achieved via amplitude and frequency modulation of the provided boundary condition (BC) spectra over space and time. To reduce the discrepancy between observations and SWAN model predictions, we rely on a correction of the prescribed boundary condition energy density, $E_{\rm BC, \ nested}$, given by
\begin{equation}
E_{\rm BC, \ corrected} (\omega, \theta, \vec{x}, t) = a_1 (t) \ E_{\rm BC, \ nested} (a_2 (t) \ \omega, \theta, \vec{x}, t)
\end{equation}
with amplitude and frequency modulation factors, $a_1 (t)$ and $a_2 (t)$, respectively. The modulation factors follow an exponential decay/growth model with time-varying decay/growth rates, parametrized by a total of four unknown parameters, as in
\begin{align}\label{eq:bc_modulation}
a_i (t) & = {\rm exp} \left( \alpha_i \ {\rm exp} \left(b_i (t) \ (t - t_{\rm ref}) \right) \right) ,  & \ & i = 1, 2 \ , \\\label{eq:bc_decay_rate}
b_i (t) & = \frac{1}{2} \left( \beta_i + \gamma_1 \right) \left( 1 - {\rm tanh} \ \left( \gamma_2 (t - t_{\rm ref} - \tau) \right) \right) - \gamma_1,  & \ & i = 1, 2 \ .
\end{align}

The unknown parameters $\alpha_1$ and $\alpha_2$ control the initial values of the modulation factors $a_1$ and $a_2$ while $\beta_1$ and $\beta_2$ determine the initial values of the corresponding decay rates $b_1$ and $b_2$ (or equivalently the initial gradient of the modulation factors). The time offset $t_{\rm ref}$ is chosen to be the last time at which data was assimilated. Therefore, for each modulation factor, we have introduced two degrees of freedom to control their initial values as well as their decay rates in time. By design, the modulation factors are strictly positive due to the first exponential operator in Eq.~(\ref{eq:bc_modulation}). Furthermore, the same factors are guaranteed to approach a value of one (i.e. no modulation) over longer time forecasts since the decay rate $b$ converges to $\gamma_2$ (a positive constant) as shown in Eq.~(\ref{eq:bc_decay_rate}). Therefore, regardless of the values of the four parameters, the corrected BC spectra converge to the pre-specified ones for large time (roughly > 4 days) regardless of what the data suggests. This is achieved by choosing $\gamma_1 = 0.03$, $\gamma_2 = 0.05$, and $\tau = 24 {\rm \ hours}$, with those values obtained using optimization experiments on simplified models. This modeling choice is based on the observation (through statistical studies, not shown for brevity) that any discrepancy in the BC spectral bulk parameters tend to persist on average for a couple of days.

\begin{figure}[ht!]
\centering
\begin{subfigure}[b]{.4\columnwidth}
  \centering
  \includegraphics[width=\columnwidth]{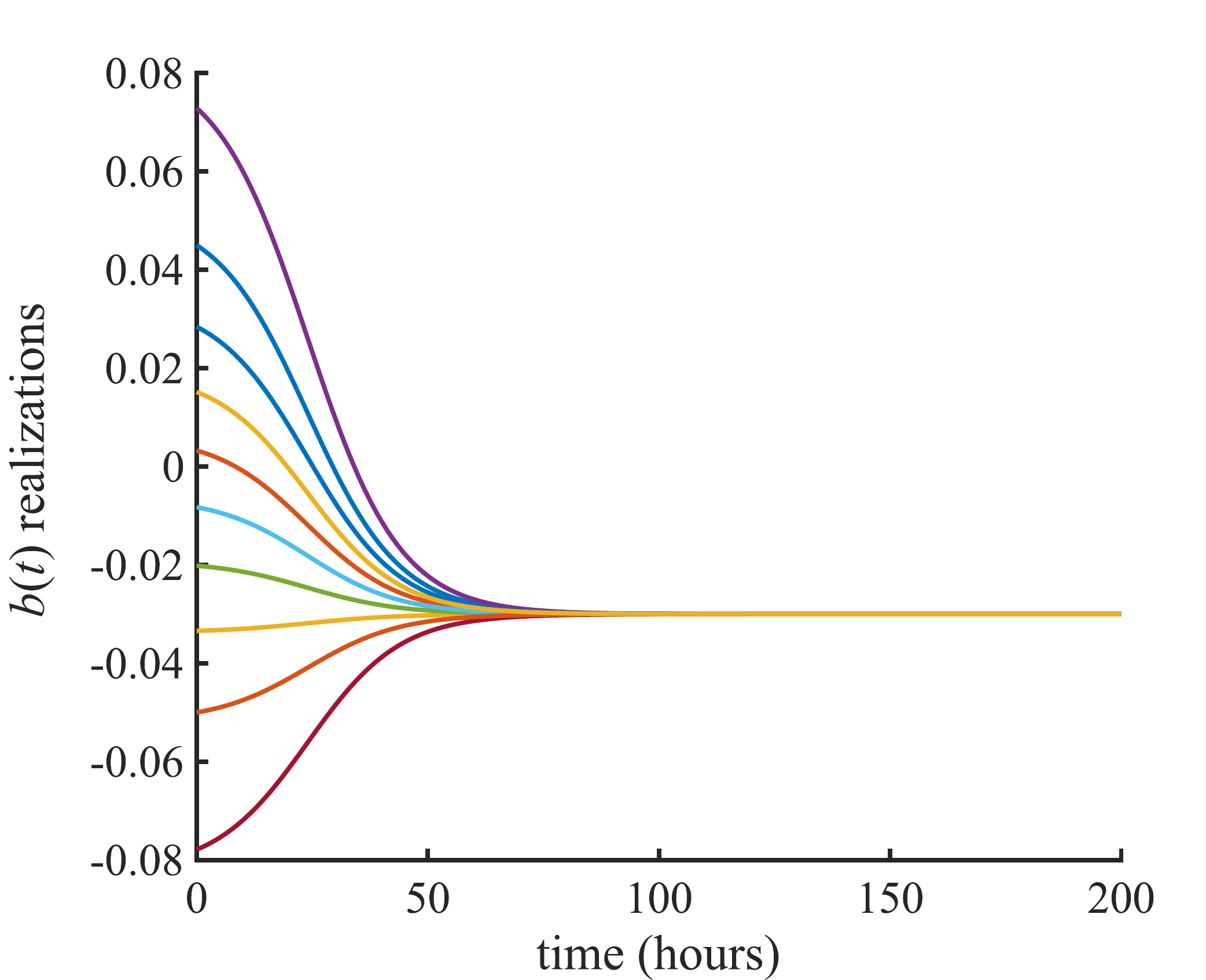}
  \caption{}
  \label{PC}
\end{subfigure}
\begin{subfigure}[b]{.4\columnwidth}
  \centering
  \includegraphics[width=\columnwidth, keepaspectratio]{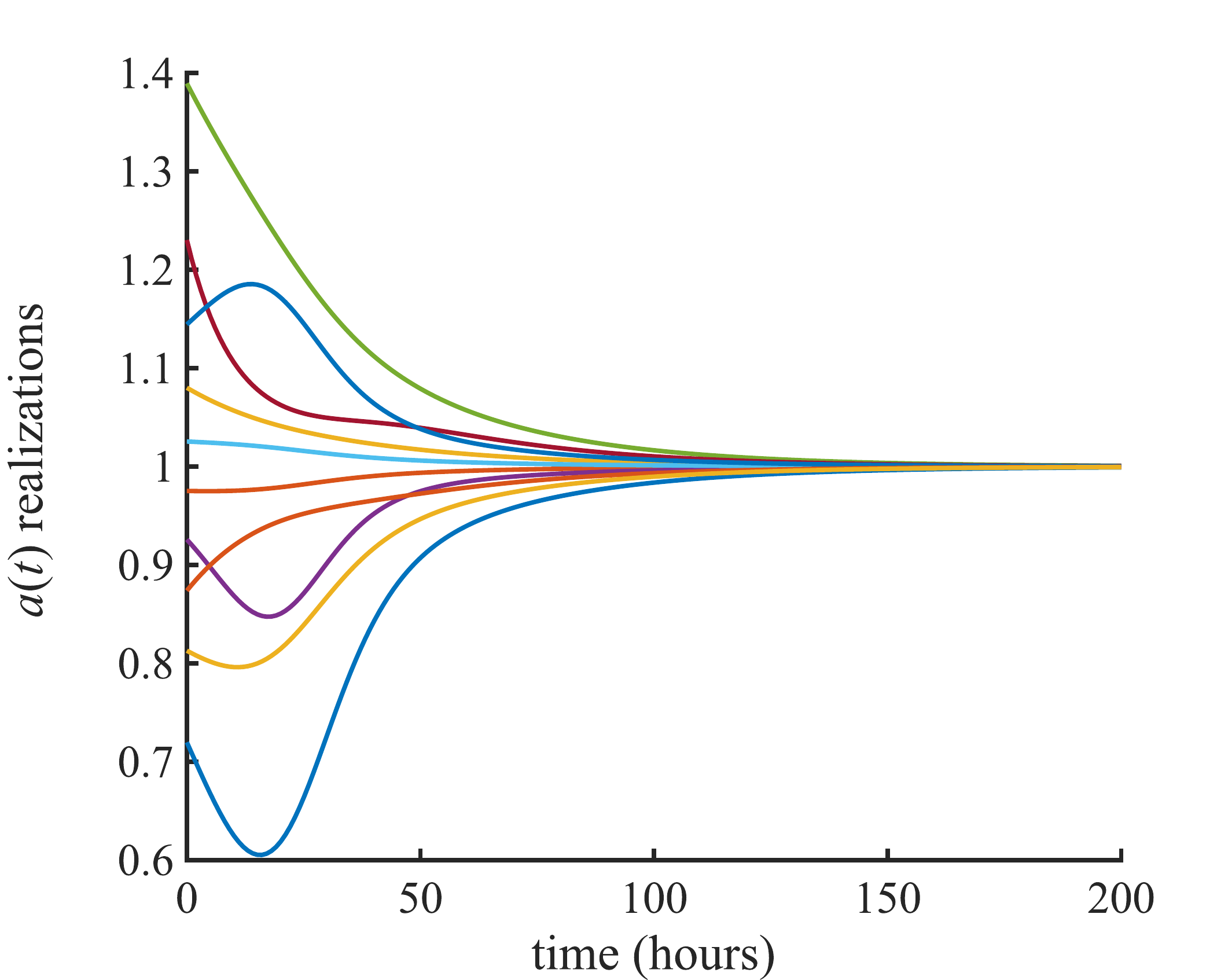}
  \caption{}
  \label{TL}
\end{subfigure}
\caption{Ten Monte Carlo realizations of the modulation factor $a(t)$ and associated decay rate $b(t)$ for parameters $\alpha$ and $\beta$ with means of zero and standard deviations of 0.2 for $\alpha$ and 0.05 for $\beta$}
\label{fig:mod_factors_1}
\end{figure}

\begin{figure}[!ht]
    \centering
    \includegraphics[width=0.4\columnwidth]{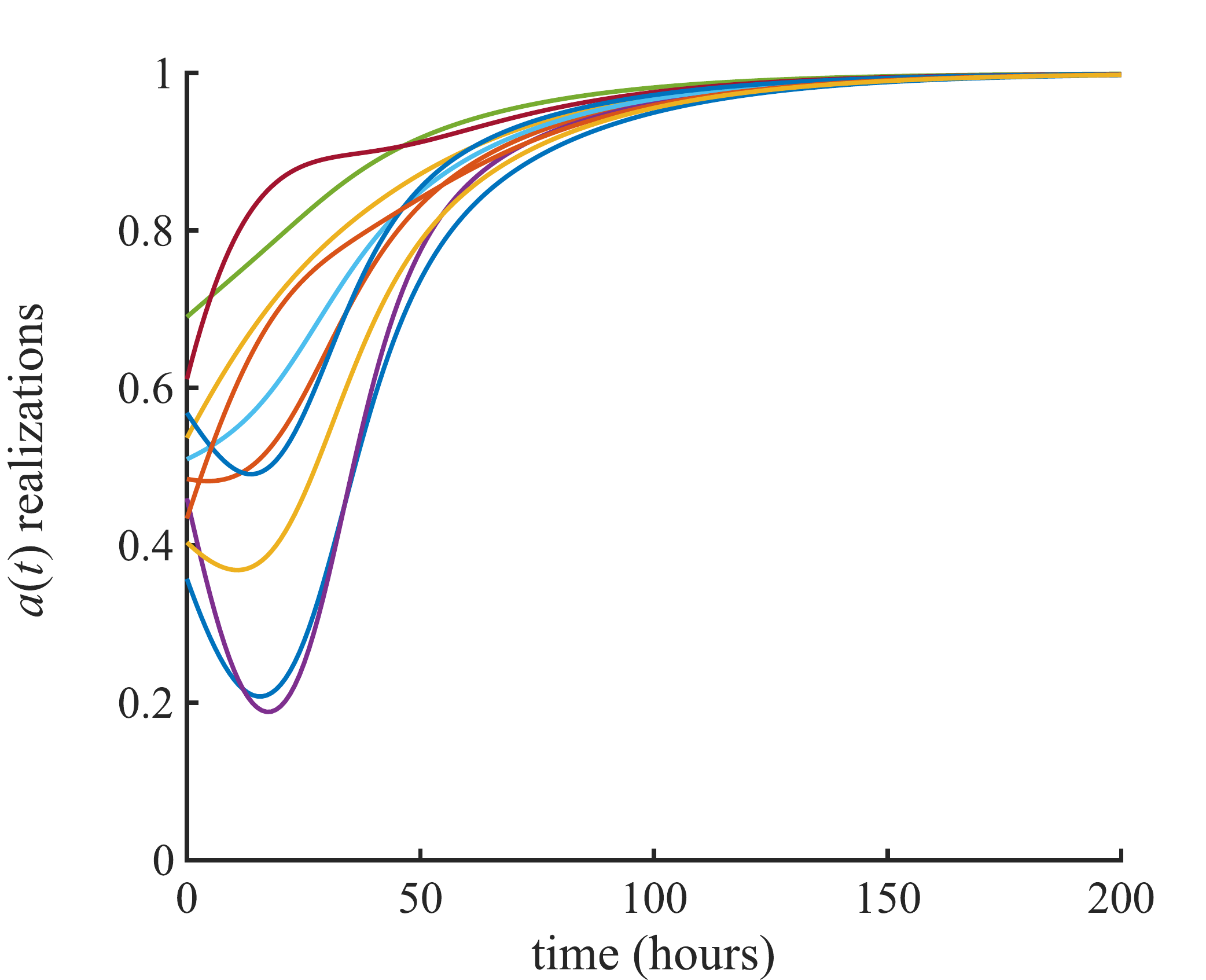}
    \caption{Ten Monte Carlo realizations of the modulation factor $a(t)$ for parameters $\alpha$ and $\beta$ with means of -0.7 for $\alpha$ and 0 for $\beta$ and standard deviations of 0.2 for $\alpha$ and 0.05 for $\beta$.}
    \label{fig:mod_factors_2}
\end{figure}

In addition to correcting the state vector, comprised of two spectral bulk parameters (significant wave height and energy period) across all nodes in the computational domain, the four parameters that control the temporal behavior of the boundary condition amplitude and frequency modulation factors will also be updated using data assimilation by augmenting the state vector with the four unknown parameters. In a data assimilation context, we are thus performing joint state and parameter estimation. Since the assimilation of data reduces the uncertainty in those parameters (and subsequently the uncertainty in BC spectra) over time, we artificially inflate the variance of the unknown parameters while maintaining their mean values (using a linear transformation) to maintain a minimum allowable level of uncertainty. This minimum allowable standard deviations, 0.2 for $\alpha_i$ and 0.05 for $\beta_i$, were chosen based on numerical experiments using synthetic data in order to cause sufficient levels of uncertainty in the state for forecast purposes. Fig.~\ref{fig:mod_factors_1} provides Monte Carlo realizations of the modulation factor $a(t)$ and associated decay rate $b(t)$ for parameters $\alpha$ and $\beta$ with Gaussian distributions of zero mean. Furthermore, Fig.~\ref{fig:mod_factors_2} provides Monte Carlo realizations of the modulation factor $a(t)$ for a Gaussian $\alpha$ parameter with a mean of -0.7 to illustrate the ability to capture an initial discrepancy in amplitude or frequency while also converging back to a modulation factor of 1 (i.e. no effective modulation). See [\cite{khalil2015estimation}] for a discussion on artificial inflation of parameter variance in data assimilation.

Strong cross-correlations across the entire state space were observed from Monte Carlo SWAN model simulations, suggesting that a small ensemble for EnKF would suffice in this context. It was determined that an ensemble of 10 to 20 model runs is sufficient to capture the statistical error characteristics in the state of the system. Therefore, we relied upon an EnKF ensemble size of 20 for all results included in this paper.

\subsection{Data Assimilation Workflows}

\begin{figure}[!ht]
    \centering
    \includegraphics[width=0.85\columnwidth]{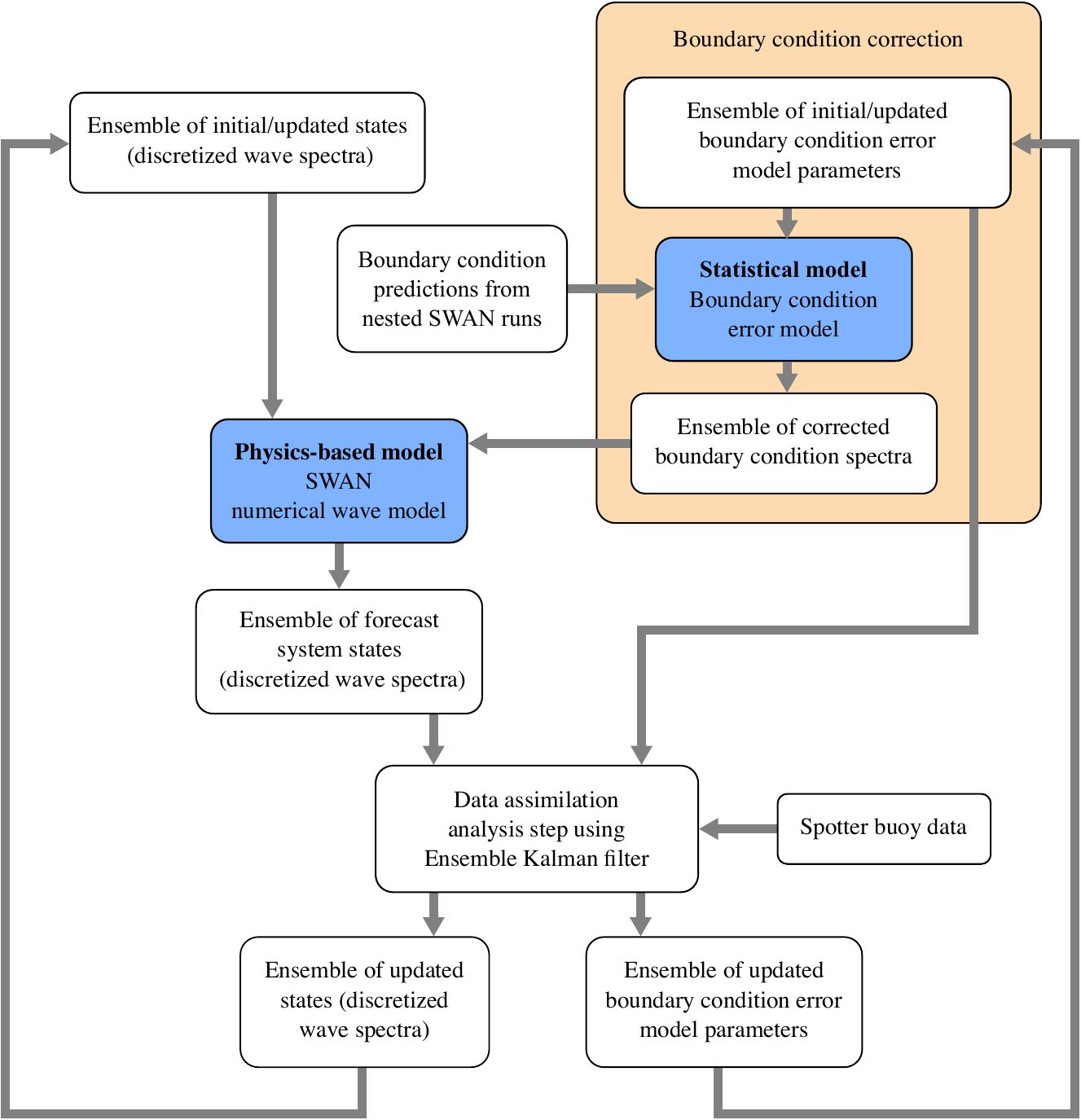}
    \caption{A schematic flowchart of the data assimilation framework as applied to the joint state and parameter inference for wave characteristic predictions.}
    \label{fig:workflows}
\end{figure}

Coupling the previously described SWAN numerical wave model for forecasting, ensemble Kalman filter for data assimilation, along with the boundary condition corrections, we obtain workflows as shown in Fig.~\ref{fig:workflows}. The flowchart illustrates the interplay between key components such as the wave and boundary condition error models, the data assimilation forecast and analysis steps, and the available buoy data. The two modeling components, colored in blue, relate to the physics-based numerical wave model and the data-driven error model used in correcting the boundary conditions available from offline simulations involving nested SWAN simulations. This hybrid modeling approach, combining physics-based and data-based modeling strategies, results in greater accuracy of numerical wave forecasts.
 
\section{Results and Discussion}\label{sec:results}

\begin{figure}[ht!]
\centering
\begin{subfigure}[b]{.4\columnwidth}
  \centering
  \includegraphics[width=\columnwidth, keepaspectratio]{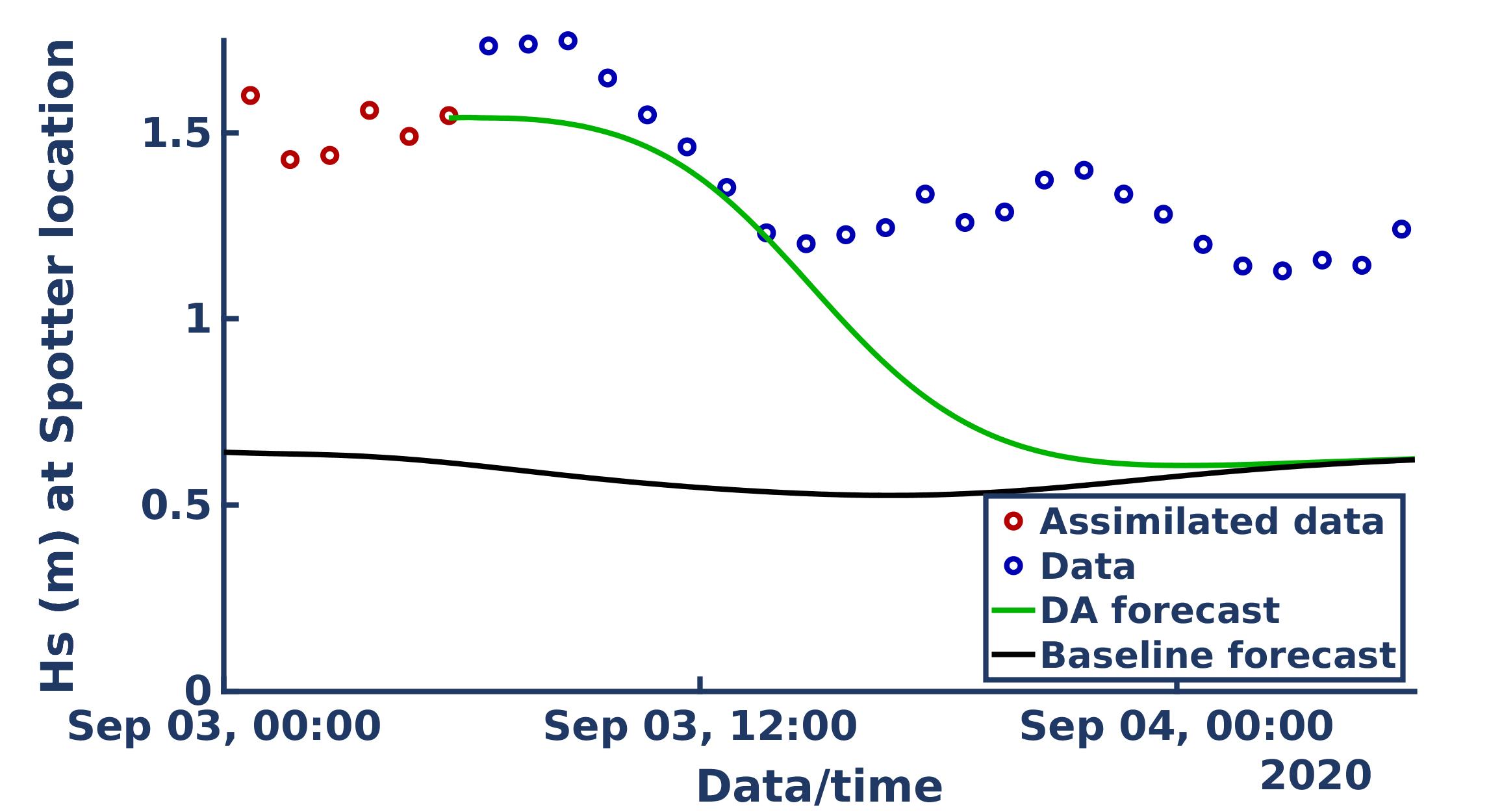}
  \caption{$H_s$}
  \label{fig:good_hs}
\end{subfigure}
\begin{subfigure}[b]{.4\columnwidth}
  \centering
  \includegraphics[width=\columnwidth, keepaspectratio]{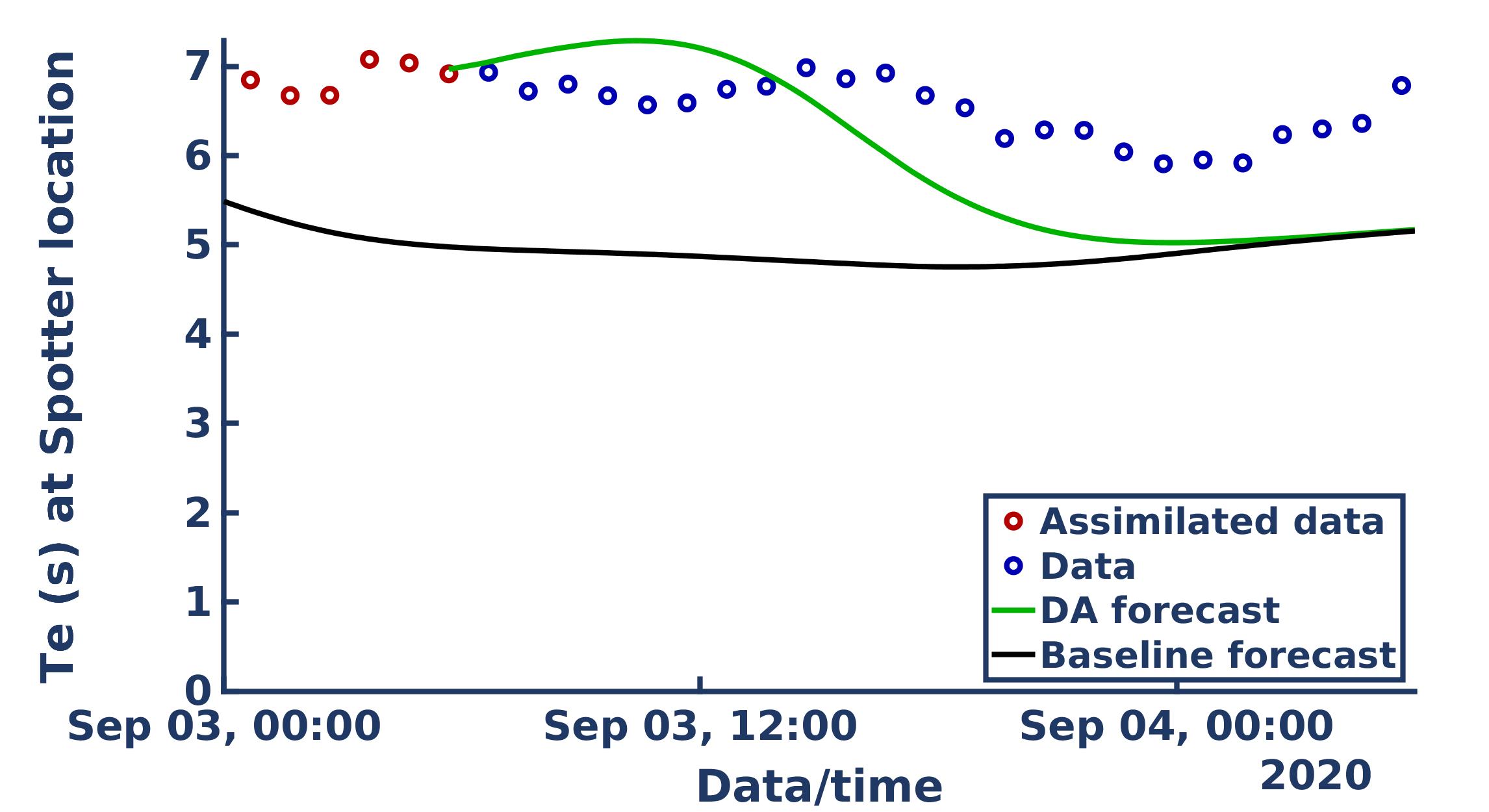}
  \caption{$T_e$}
  \label{fig:good_te}
\end{subfigure}
\caption{Single data-assimilation experiment for forecasting starting at 6am on September 3. Both $H_s$ and $T_e$ are shown at the measurement (Spotter 0397 at the boundary).}
\label{fig:good}
\end{figure}

A series of data assimilation experiments was performed using the Spotter buoy data from Yakutat, Alaska for the period of May 18 - September 23, 2020. 
For each experiment, first the nested SWAN runs described above were ran to obtain the baseline computational predictions. 
Then, EnKF was used to perform data assimilation for a six hours period, with data coming in about once per hour, followed by a 24 hour forecast starting from the corrected state. 
Six hours were found to be sufficient to mimic a long running data-assimilation scheme. The goal is to compare the baseline predictions to the forecasts into the future using data-assimilation. 
This forecast would be used to inform grid operations. 
The baseline prediction is based on the SWAN model driven by Wave Watch III conditions, which do incorporate wind data assimilation, but not the local data from the spotter buoy. 
The data from Spotter 0397 at the boundary was used for data assimilation, while Spotter 0107 is considered the location of interest, i.e., where a \ac{WEC} would be located. 
An example of one of these data assimilation experiments is shown in Fig.~\ref{fig:good}. 
It can be seen that the data-assimilated predictions start from the corrected state, follow the data trend for some period of time, and eventually return to the baseline prediction. 
This is the desired behavior enforced through the boundary condition model described above. 

\begin{figure}[ht!]
\centering
\begin{subfigure}[b]{.4\columnwidth}
  \centering
  \includegraphics[width=\columnwidth, keepaspectratio]{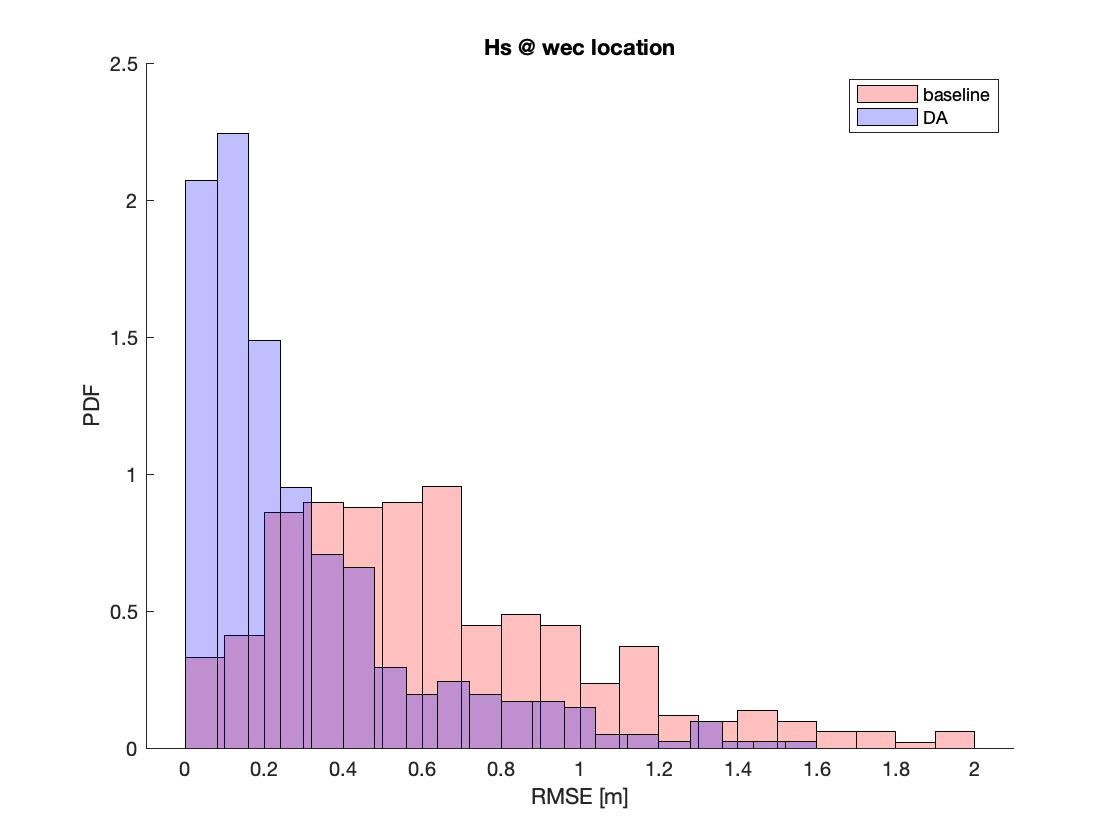}
  \caption{$H_s$}
  \label{fig:sep_hs}
\end{subfigure}
\begin{subfigure}[b]{.4\columnwidth}
  \centering
  \includegraphics[width=\columnwidth, keepaspectratio]{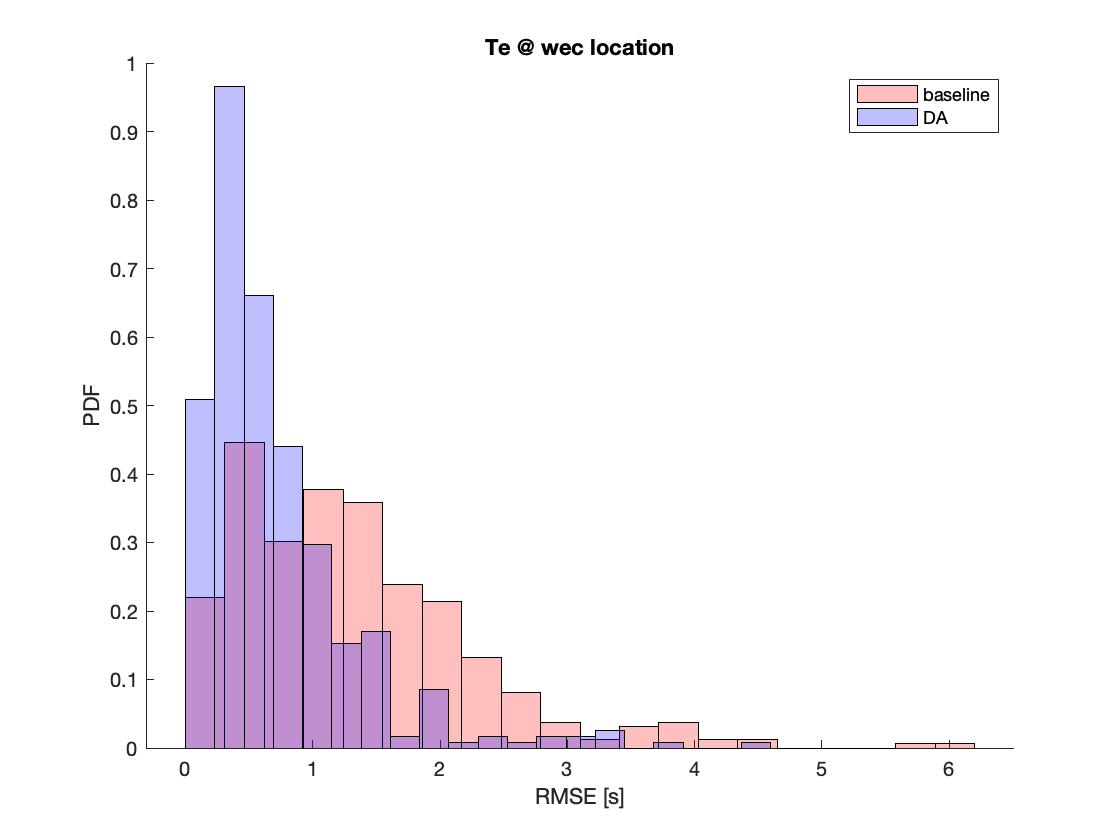}
  \caption{$T_e$}
  \label{fig:sep_te}
\end{subfigure}
\caption{Comparison of predictive error (RMSE) in the forecast between the baseline and data-assimilation results at the proposed \ac{WEC} location (Spotter 0107).}
\label{fig:sep}
\end{figure}

Similar experiments where performed every six hours for the period from May 18 - September 23, 2020, after removing missing (e.g. wind data was missing for several days in this period) or corrupt data. 
A total of 403 experiments were performed. 
We compare the 6-hour forecast at the location of the \ac{WEC}, i.e., where Spotter 0107 (not used for data-assimilation) is located. 
Fig.~\ref{fig:sep} shows the RMSE for both the baseline and data-assimilation predictions for all experiments. 
The RMSE is calculated based on the norm of the error vector for the 6-hour period (i.e., the difference between prediction and measurement for all spotter observations in that period) as 
\begin{equation}
    RMSE = \frac{\lVert p-d \rVert}{\sqrt{N}},
\end{equation}
where $p$ and $d$ are the vectors of predictions and observation data, respectively, and $N$ is the number of observations (length of $p$ and $d$). 
It can be seen in Fig.~\ref{fig:sep} that the data assimilation significantly reduces the error in both $H_s$ and $T_e$. 
Table~\ref{tab:results} gives some summary statistics of the distributions shown in Fig.~\ref{fig:sep}. 

\begin{table}[ht!]
    \centering
    \begin{tabular}{l|c|c}
         \ & baseline & with data assimilation\\
         \hline
         mean $H_s$ RMSE & 0.627 m & 0.292 m\\
         scatter (standard deviation) in $H_s$ RMSE & 0.390 m & 0.290 m \\
         mean $T_e$ RMSE & 1.32 s & 0.711 s\\
         scatter (standard deviation) in $T_e$ RMSE  & 0.924 s & 0.615 s\\
    \end{tabular}
    \caption{Summary statistics for the predictive error (RMSE) from all experiments.}
    \label{tab:results}
\end{table}

\begin{figure}[ht!]
\centering
\begin{subfigure}[b]{.4\columnwidth}
  \centering
  \includegraphics[width=\columnwidth, keepaspectratio]{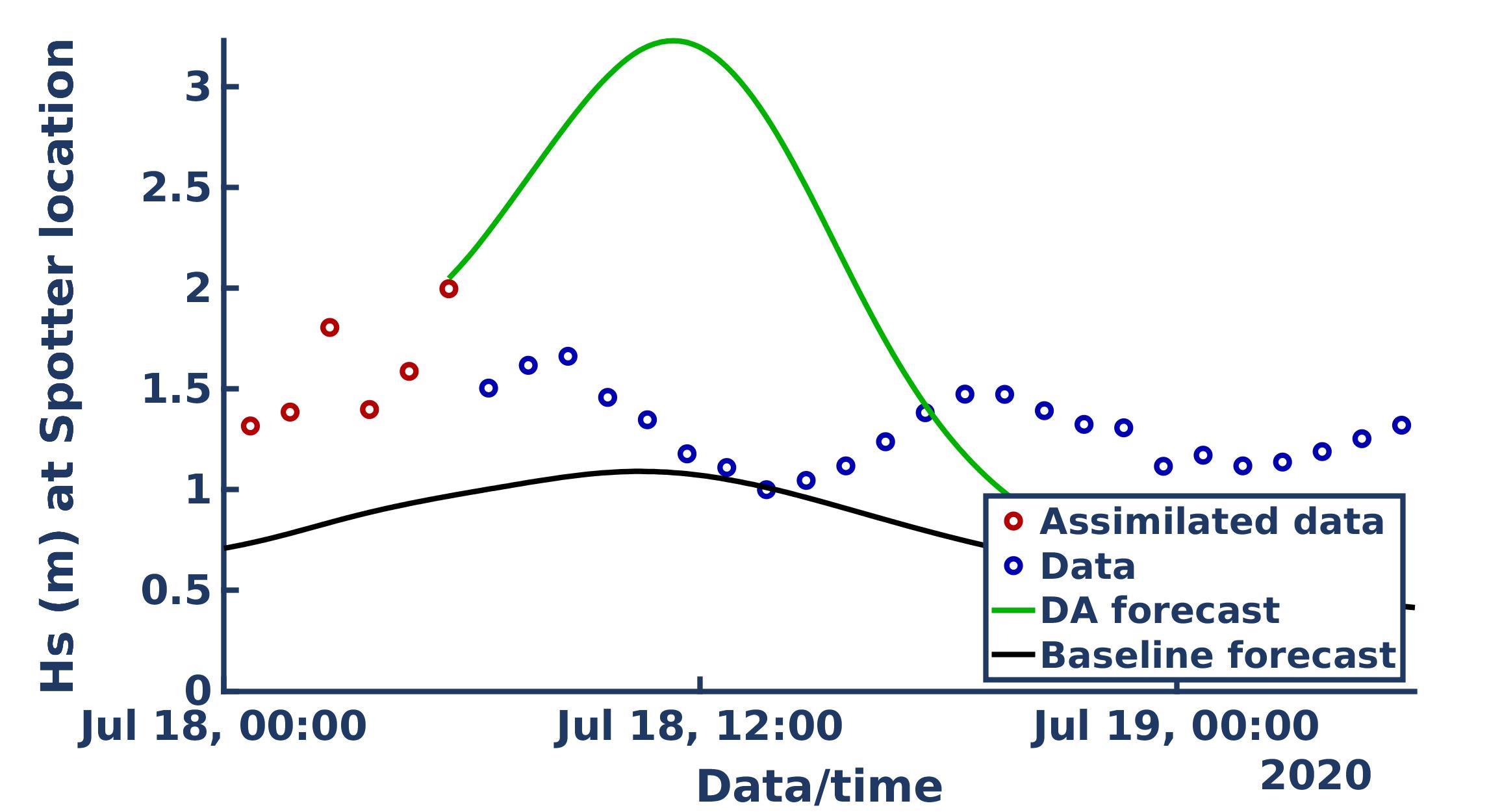}
  \caption{$H_s$}
  \label{fig:bad_hs}
\end{subfigure}
\begin{subfigure}[b]{.4\columnwidth}
  \centering
  \includegraphics[width=\columnwidth, keepaspectratio]{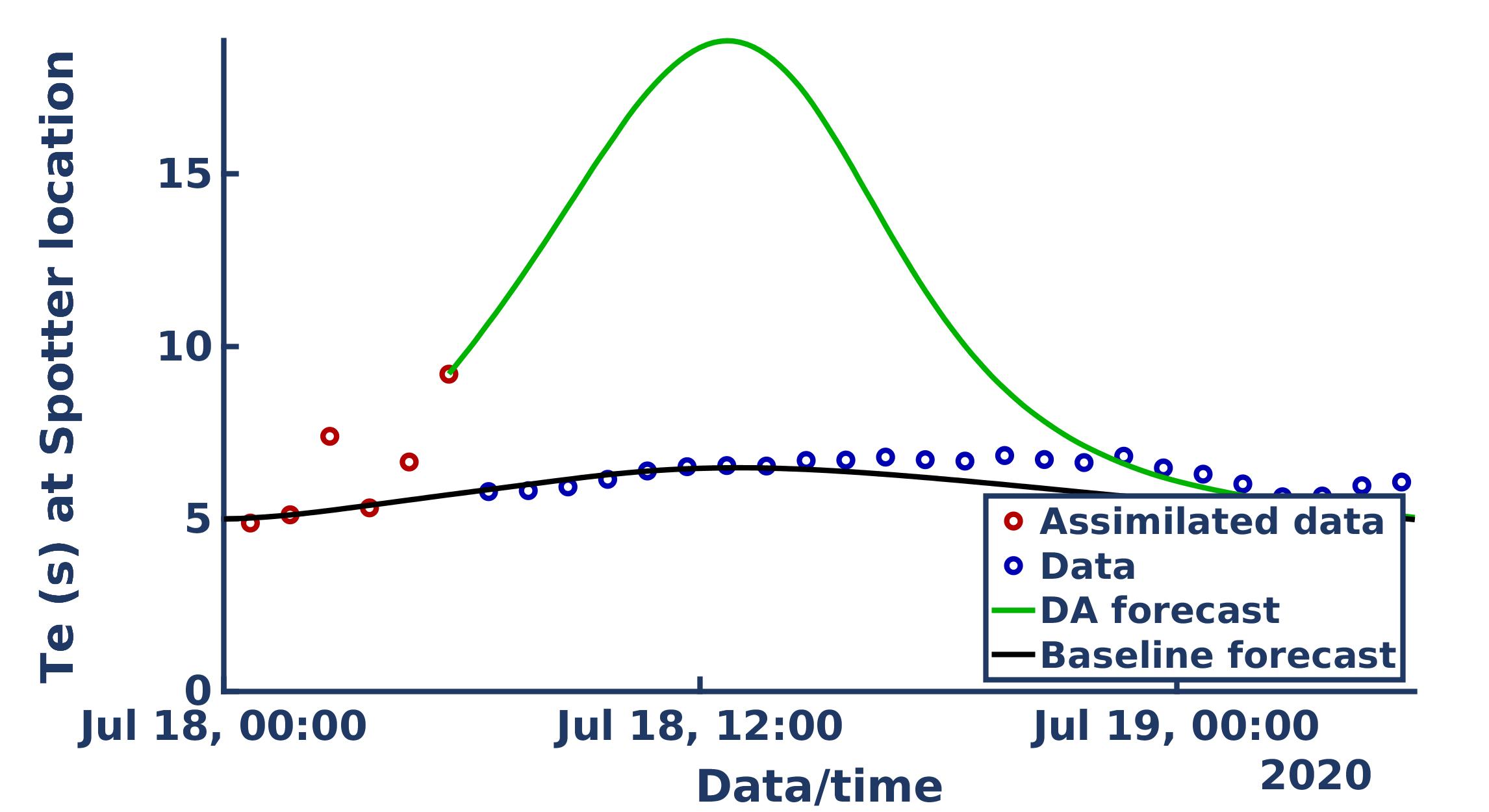}
  \caption{$T_e$}
  \label{fig:bad_te}
\end{subfigure}
\caption{Single data-assimilation experiment for forecasting starting at 6am on July 18. Both $H_s$ and $T_e$ are shown at the measurement (Spotter 0397 at the boundary).}
\label{fig:bad}
\end{figure}

\begin{figure}[ht!]
\centering
\begin{subfigure}[b]{.4\columnwidth}
  \centering
  \includegraphics[width=\columnwidth, keepaspectratio]{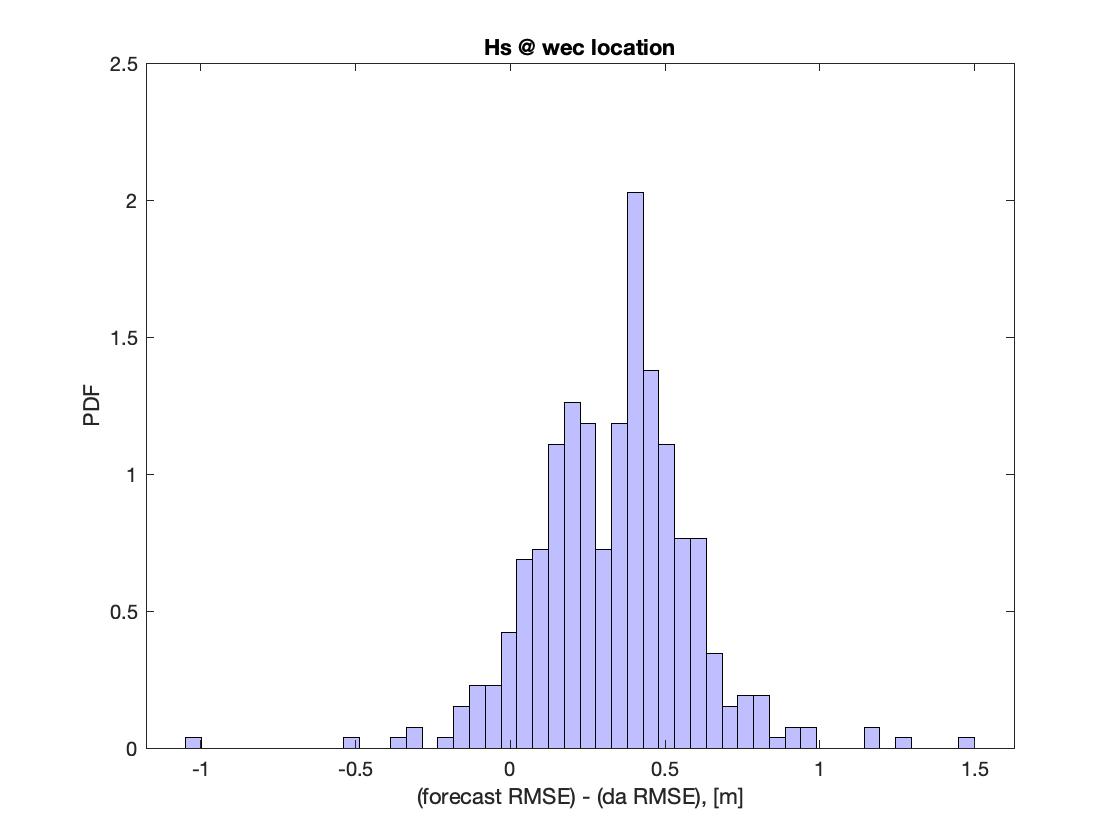}
  \caption{$H_s$}
  \label{fig:rel_hs}
\end{subfigure}
\begin{subfigure}[b]{.4\columnwidth}
  \centering
  \includegraphics[width=\columnwidth, keepaspectratio]{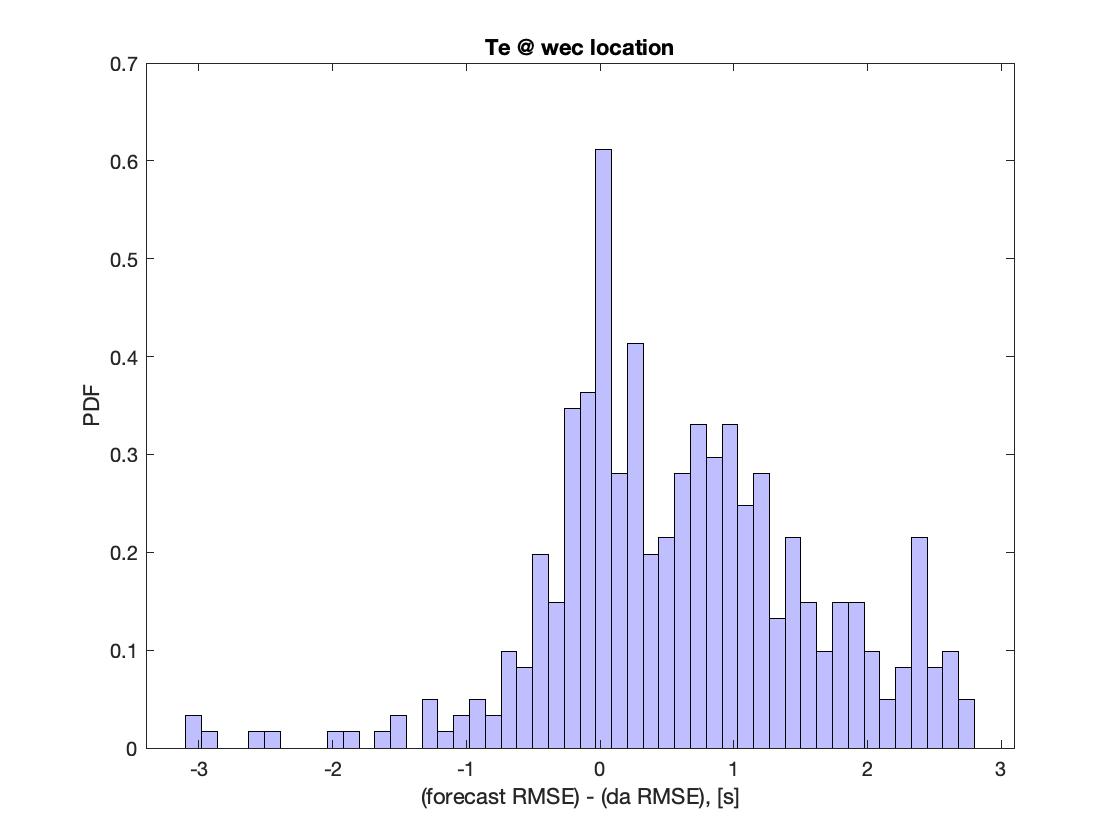}
  \caption{$T_e$}
  \label{fig:rel_te}
\end{subfigure}
\caption{Relative error between the data-assimilation and baseline forecasts at the \ac{WEC} location (Spotter 0107).}
\label{fig:rel}
\end{figure}

Although the implemented data-assimilation framework works as expected it can occasionally result in larger errors than the baseline. 
As an example, Fig.~\ref{fig:bad} shows a particularly bad case where the trend of the data changes significantly after data-assimilation is ended. 
The predictions are seen to follow the trend of the last 3 observations which would suggest the $H_s$ and $T_e$ are very rapidly increasing. 
The resulting predicted sea-states are nonphysical (e.g., $H_s$ of more than 15m).  
To address these cases we capped the data-assimilation-based forecasting to twice the baseline prediction for $H_s$ and 1.5 times the baseline prediction for $T_e$. 
The results shown in Fig.~\ref{fig:sep} and Fig.~\ref{fig:rel} use this capping. 
In future work, a more rigorous regularization technique could be used. 
Fig.~\ref{fig:rel} shows the results of all experiments (same results as in Fig.~\ref{fig:sep}) as a relative error, i.e., the difference between the baseline and data-assimilation RMSE. 
While most values are positive---indicating an error reduction when using data-assimilation---it can be seen that for some of the experiments the data-assimilation resulted in increased error. 
This is more common for the $T_e$ predictions. 
A more rigorous regularization should help reduce both the frequency and severity of these cases. 

\section{Conclusion}\label{sec:conclusion}

In this paper we presented the development and assessment of a wave modeling framework with real-time data assimilation capabilities for wave energy converter power prediction. Available real-time wave spectra from low-cost wave measurement buoys allowed for operational data assimilation with the ensemble Kalman filter technique within a hybrid modeling procedure whereby physics-based numerical wave models are combined with data-driven error models that aim to capture the discrepancy in prescribed boundary conditions. The measured wave spectra are assimilated for combined state and parameter estimation while taking into account model and observational errors. The analysis of incoming data allowed for more accurate and precise wave characteristic predictions at the locations of interest. Initial deployment data obtained offshore Yakutat, Alaska, indicated that measured wave data from one buoy that were assimilated into the wave modeling framework resulted in improved forecast skill in comparison to traditional numerical forecasts.

\section*{Acknowledgments}
Chris Flanary at Integral Consulting, Inc., set up and ran the WaveWatchIII model, which provided boundary conditions into the coarse SWAN domain.
Sandia National Laboratories is a multimission laboratory managed and operated by National Technology and Engineering Solutions of Sandia, LLC., a wholly owned subsidiary of Honeywell International, Inc., for the U.S. Department of Energy's National Nuclear Security Administration under contract DE-NA-0003525.
The views expressed in the article do not necessarily represent the views of the U.S. Department of Energy or the United States Government.

\appendix
\section{Updating of Wave Energy Density to Macth Spectral Parameters} \label{app:spec_update}
Herein, we provide supplementary details to describe a particular transformation of 2D wave energy spectra with resulting bulk parameters matching target (pre-specified) values that are provided from the analysis step of EnKF. In particular, we will focus on two bulk parameters, namely the significant wave height and mean energy period. We will show how the transformation parameters can be obtained analytically using the starting and target bulk parameter values. 

We start with a 2D wave energy spectrum, $E \left( \omega, \theta \right)$, with the associated relevant spectral moments
\begin{align}
m_0 & = \int E \left( \omega, \theta \right) {\rm d} \omega \, {\rm d} \theta \ ,\\
m_{-1} & = \int \omega^{-1} E \left( \omega, \theta \right) {\rm d} \omega \, {\rm d} \theta \ ,
\end{align}
and relelavnt bulk parameters, namely significant wave height, $H_s$, and mean energy period, $T_e$, given by
\begin{align}
H_s & = 4 \sqrt{m_0} \ ,\\
T_e & = \frac{m_{-1}}{m_0}\ .
\end{align}

This spectrum might be describing the wave characteristics at a particular computational node, whether along the boundary or inside the domain of interest. In data assimilation workflows, we will often face the task of updating the wave spectra to match a "target" significant wave height and mean energy period. Let's denote those by $\hat H_s$ and $\hat T_e$, respectively. We will use the following spectral transformation \footnote{There are other transformations that one might use. This particular choice guarantees physically valid spectra as long as the target bulk parameters are themselves physically valid.} to arrive at those target values:
\begin{align}\label{eq:transformation}
\hat{E} \left( \omega, \theta \right) = a_1 \, E \left( a_2 \, \omega, \theta \right) \ .
\end{align}

It can shown that the above transformation results in new spectral moments that can be related to the original one analytically using
\begin{align}
\hat m_0 & = a_1 \, a_2 \, m_0 \ ,\\
\hat m_{-1} & = \frac{a_1}{a_2} \, m_{-1} \ .
\end{align}

The above expression can be used to derive the new bulk parameters that correspond to the transformed spectrum:

\begin{align}
\hat H_s & = 4 \, \sqrt{\hat m_0} \nonumber \\
& = \sqrt{a_1 \, a_2} \, 4  \, \sqrt{m_0} \nonumber \\
& = \sqrt{a_1 \, a_2} \,  H_s \ , \label{eq:bulk_identity_1}
\end{align}
and
\begin{align}
\hat T_e & = \frac{\hat m_{-1}}{\hat m_0} \nonumber \\
& = \frac{\frac{a_1}{a_2} m_{-1}}{a_1 \, a_2 \, m_0} \nonumber \\
& = \frac{1}{a_2^2} \frac{m_{-1}}{m_0} \nonumber \\
& = \frac{1}{a_2^2} T_e \ . \label{eq:bulk_identity_2}
\end{align}

Equations (\ref{eq:bulk_identity_1}) and (\ref{eq:bulk_identity_2}) can be used to solve for the two unknown coefficients that are used in transforming the spectrum (see Equation (\ref{eq:transformation})), resulting in the following amplitude and frequency modulation to achieve the target bulk parameter values.
\begin{align}
a_1 & = \left( \frac{\hat H_s}{H_s} \right)^2  \frac{\hat T_e }{T_e } \ ,\\
a_2 & = \sqrt{\frac{T_e }{\hat T_e }}  \ .
\end{align}

\bibliographystyle{cas-model2-names}


\begin{thebibliography}{17}
\expandafter\ifx\csname natexlab\endcsname\relax\def\natexlab#1{#1}\fi
\providecommand{\url}[1]{\texttt{#1}}
\providecommand{\href}[2]{#2}
\providecommand{\path}[1]{#1}
\providecommand{\DOIprefix}{doi:}
\providecommand{\ArXivprefix}{arXiv:}
\providecommand{\URLprefix}{URL: }
\providecommand{\Pubmedprefix}{pmid:}
\providecommand{\doi}[1]{\href{http://dx.doi.org/#1}{\path{#1}}}
\providecommand{\Pubmed}[1]{\href{pmid:#1}{\path{#1}}}
\providecommand{\bibinfo}[2]{#2}
\ifx\xfnm\relax \def\xfnm[#1]{\unskip,\space#1}\fi
\bibitem[{Andrieu and Doucet(2003)}]{andrieu03}
\bibinfo{author}{Andrieu, C.}, \bibinfo{author}{Doucet, A.},
  \bibinfo{year}{2003}.
\newblock \bibinfo{title}{Online expectation-maximization type algorithms for
  parameter estimation in general state space models}, in:
  \bibinfo{booktitle}{Proceedings of ICASSP '03, the IEEE International
  Conference on Acoustics, Speech, and Signal Processing},
  \bibinfo{address}{Hong Kong}.
\bibitem[{Booij et~al.(1999)Booij, Ris and Holthuijsen}]{booij1999third}
\bibinfo{author}{Booij, N.}, \bibinfo{author}{Ris, R.C.},
  \bibinfo{author}{Holthuijsen, L.H.}, \bibinfo{year}{1999}.
\newblock \bibinfo{title}{A third-generation wave model for coastal regions: 1.
  model description and validation}.
\newblock \bibinfo{journal}{Journal of geophysical research: Oceans}
  \bibinfo{volume}{104}, \bibinfo{pages}{7649--7666}.
\bibitem[{Chui and Chen(1999)}]{chui99}
\bibinfo{author}{Chui, C.K.}, \bibinfo{author}{Chen, G.}, \bibinfo{year}{1999}.
\newblock \bibinfo{title}{Kalman Filtering with Real-time Applications}.
\newblock \bibinfo{edition}{third} ed., \bibinfo{publisher}{Springer},
  \bibinfo{address}{Berlin}.
\bibitem[{Doucet et~al.(2000)Doucet, Godsill and Andrieu}]{doucet00}
\bibinfo{author}{Doucet, A.}, \bibinfo{author}{Godsill, S.J.},
  \bibinfo{author}{Andrieu, C.}, \bibinfo{year}{2000}.
\newblock \bibinfo{title}{On sequential {Monte Carlo} sampling methods for
  {Bayesian} filtering}.
\newblock \bibinfo{journal}{Statistics and Computing} \bibinfo{volume}{10},
  \bibinfo{pages}{197--208}.
\bibitem[{Evensen(1994)}]{evensen1994sequential}
\bibinfo{author}{Evensen, G.}, \bibinfo{year}{1994}.
\newblock \bibinfo{title}{Sequential data assimilation with a nonlinear
  quasi-geostrophic model using monte carlo methods to forecast error
  statistics}.
\newblock \bibinfo{journal}{Journal of Geophysical Research: Oceans}
  \bibinfo{volume}{99}, \bibinfo{pages}{10143--10162}.
\bibitem[{Evensen(2006)}]{evensen06}
\bibinfo{author}{Evensen, G.}, \bibinfo{year}{2006}.
\newblock \bibinfo{title}{Data Assimilation: The Ensemble Kalman Filter}.
\newblock \bibinfo{publisher}{Springer}, \bibinfo{address}{Berlin}.
\bibitem[{Evensen et~al.(2009)}]{evensen2009data}
\bibinfo{author}{Evensen, G.}, et~al., \bibinfo{year}{2009}.
\newblock \bibinfo{title}{Data assimilation: the ensemble Kalman filter}.
  volume~\bibinfo{volume}{2}.
\newblock \bibinfo{publisher}{Springer}.
\bibitem[{Guedes~Soares et~al.(2011)Guedes~Soares, Rusu, Bernardino and
  Pilar}]{guedes2011operational}
\bibinfo{author}{Guedes~Soares, C.}, \bibinfo{author}{Rusu, L.},
  \bibinfo{author}{Bernardino, M.}, \bibinfo{author}{Pilar, P.},
  \bibinfo{year}{2011}.
\newblock \bibinfo{title}{An operational wave forecasting system for the
  portuguese continental coastal area}.
\newblock \bibinfo{journal}{Journal of Operational Oceanography}
  \bibinfo{volume}{4}, \bibinfo{pages}{17--27}.
\bibitem[{Kalman(1960)}]{kalman60}
\bibinfo{author}{Kalman, R.E.}, \bibinfo{year}{1960}.
\newblock \bibinfo{title}{A new approach to linear filtering and prediction
  problems}.
\newblock \bibinfo{journal}{Journal of Basic Engineering} \bibinfo{volume}{82},
  \bibinfo{pages}{35--45}.
\bibitem[{Khalil et~al.(2009)Khalil, Sarkar and Adhikari}]{khalil07_2}
\bibinfo{author}{Khalil, M.}, \bibinfo{author}{Sarkar, A.},
  \bibinfo{author}{Adhikari, S.}, \bibinfo{year}{2009}.
\newblock \bibinfo{title}{Nonlinear filters for chaotic oscillatory systems}.
\newblock \bibinfo{journal}{Journal of Nonlinear Dynamics}
  \bibinfo{volume}{55}, \bibinfo{pages}{113--137}.
\bibitem[{Khalil et~al.(2015)Khalil, Sarkar, Adhikari and
  Poirel}]{khalil2015estimation}
\bibinfo{author}{Khalil, M.}, \bibinfo{author}{Sarkar, A.},
  \bibinfo{author}{Adhikari, S.}, \bibinfo{author}{Poirel, D.},
  \bibinfo{year}{2015}.
\newblock \bibinfo{title}{The estimation of time-invariant parameters of noisy
  nonlinear oscillatory systems}.
\newblock \bibinfo{journal}{Journal of Sound and Vibration}
  \bibinfo{volume}{344}, \bibinfo{pages}{81--100}.
\bibitem[{Kitagawa(1998)}]{kitagawa98}
\bibinfo{author}{Kitagawa, G.}, \bibinfo{year}{1998}.
\newblock \bibinfo{title}{A self-organizing state-space model}.
\newblock \bibinfo{journal}{Journal of the American Statistical Association}
  \bibinfo{volume}{93}, \bibinfo{pages}{1203--1215}.
\bibitem[{Komen et~al.(1996)Komen, Cavaleri, Donelan, Hasselmann, Hasselmann
  and Janssen}]{komen1996dynamics}
\bibinfo{author}{Komen, G.J.}, \bibinfo{author}{Cavaleri, L.},
  \bibinfo{author}{Donelan, M.}, \bibinfo{author}{Hasselmann, K.},
  \bibinfo{author}{Hasselmann, S.}, \bibinfo{author}{Janssen, P.},
  \bibinfo{year}{1996}.
\newblock \bibinfo{title}{Dynamics and modelling of ocean waves}.
\bibitem[{Ristic et~al.(2004)Ristic, Arulampalam and Gordon}]{ristic04}
\bibinfo{author}{Ristic, B.}, \bibinfo{author}{Arulampalam, S.},
  \bibinfo{author}{Gordon, N.}, \bibinfo{year}{2004}.
\newblock \bibinfo{title}{Beyond the Kalman Filter: Particle Filters for
  Tracking Applications}.
\newblock \bibinfo{publisher}{Artech House}, \bibinfo{address}{Boston}.
\bibitem[{Rusu and Raileanu(2016)}]{rusu2016multi}
\bibinfo{author}{Rusu, E.}, \bibinfo{author}{Raileanu, A.},
  \bibinfo{year}{2016}.
\newblock \bibinfo{title}{A multi-parameter data-assimilation approach for wave
  prediction in coastal areas}.
\newblock \bibinfo{journal}{Journal of Operational Oceanography}
  \bibinfo{volume}{9}, \bibinfo{pages}{13--25}.
\bibitem[{Tanizaki(1996)}]{tanizaki96}
\bibinfo{author}{Tanizaki, H.}, \bibinfo{year}{1996}.
\newblock \bibinfo{title}{Nonlinear Filters: Estimation and Applications}.
\newblock \bibinfo{edition}{second} ed., \bibinfo{publisher}{Springer},
  \bibinfo{address}{Berlin}.
\bibitem[{Tschetter et~al.(2016)Tschetter, Kasper and
  Duvoy}]{tschetter2016yakutat}
\bibinfo{author}{Tschetter, T.}, \bibinfo{author}{Kasper, J.},
  \bibinfo{author}{Duvoy, P.}, \bibinfo{year}{2016}.
\newblock \bibinfo{title}{Yakutat area wave resource assessment}.
\newblock \bibinfo{journal}{Alaska Center for Energy and Power, Alaska Center
  for Energy and Power} .

\end{thebibliography}

\end{document}